\documentclass[twocolumn,showpacs,preprintnumbers,amsmath,amssymb]{revtex4}
\usepackage{graphicx}% Include figure files
\usepackage{dcolumn}% Align table columns on decimal point
\usepackage{bm}% bold math

\begin{document}
\title{ Low-momentum ring diagrams  \\
 of neutron matter at and near the unitary limit}
\author{L.-W.\ Siu, T.\ T.\ S.\ Kuo}
%\email{thomas.kuo@stonybrook.edu}
\affiliation{Department of Physics and Astronomy,
Stony Brook University, NY 11794-3800, USA}
\author{R. Machleidt}
\affiliation{Department of Physics and Astronomy,
University of Idaho, Moscow, ID 83844, USA}

\begin{abstract}
We study neutron matter at and near the
unitary limit  using a low-momentum ring diagram approach.
By slightly tuning the meson-exchange CD-Bonn potential,
neutron-neutron potentials with various $^1S_0$ scattering
lengths such as $a_s=-12070fm$ and $+21fm$ are constructed.
 Such potentials
are renormalized with rigorous procedures to give the
corresponding $a_s$-equivalent low-momentum potentials $V_{low-k}$
, with which the low-momentum particle-particle hole-hole
ring diagrams are summed up to all orders, giving the
ground state energy $E_0$ of neutron matter for various
scattering lengths.
At the limit of $a_s\rightarrow \pm \infty$, our calculated ratio of $E_0$
to that of the non-interacting case is
found remarkably close to a constant of 0.44 over a wide range
of Fermi-momenta.  This
result reveals an universality that
is well consistent with the recent experimental and Monte-Carlo
computational study on low-density cold Fermi gas at the unitary limit.
The overall behavior of this ratio obtained with various scattering lengths
 is presented and discussed. Ring-diagram results obtained with $V_{low-k}$
 and those with $G$-matrix interactions are compared.

%By integrating out the
%high-momentum components beyond a momentum
%scale $\Lambda$ $(\sim 2 fm^{-1})$, $a_s$-equivalent neutron
%low-momentum potentials are derived.
%Using the above low-momentum interactions,
%the  ground-state energy $E_0$
%of neutron matter is calculated by summing up the low-momentum ($<\Lambda$)
%particle-particle hole-hole ring
%diagrams to all orders .
%The dependence of our results on $k_F$, the Fermi momentum, and on $a_s$
%is discussed. At the unitary limit ($a_s \rightarrow \pm \infty$ ),
%the ratio $\xi \equiv E_0/E_0^{free}$, the latter being the
%non-interacting ground-state  energy, given by our ring-diagram calculations
%is remarkably close to 0.44.
%Calculations  using self-consistent Brueckner-Hartree-Fock
%theory have also been performed, obtaining $\xi \approx 0.49$.

\end{abstract}
\pacs{pacs} \maketitle
\section{Introduction}
Back in 1999, Bertsch\cite{bishop01} formulated a many-body problem, asking:
what are the ground state properties of a two-species fermion system that has
a zero-range interaction and an infinite scattering length? Such problem
was originally set up as a parameter-free model for a fictitious neutron matter.
Recently, as the experiments on trapped cold alkali gas
undergo huge breakthroughs,
degenerate Fermi gas with a tunable scattering length (including
$\pm \infty$) becomes accessible in laboratories\cite{fbre}.
Since then cold Fermi systems
have aroused growing attention.

The term `unitary limit' has been used by many authors to
refer to the special scenario in a low-density two-species
many-body system where
the scattering length between particles
approaches infinity. More specifically,
at the unitary limit,
the scattering length $a_s$,
the Fermi momentum $k_F$, and
the range of the interaction $r_{\mbox{int}}$
satisfy
$ |a_s|>>k_F^{-1}>> r_{\mbox{int}}$.
Under such condition, atoms are `strongly interacting', and a full
theoretical description of their properties is a challenging task in
many-body theory.
Universal behavior is expected to show up
in various aspects, including
ground state properties as discussed below,
collective excitations \cite{strin04,bulgac05,hei04,
 kinast04,kinast04b,altmeyer07,wright07}, and thermodynamic properties
 \cite{bulgac05b,bulgac06,bulgac07,kinast05,thomas05}.
Such universality can be naively understood as the `dropping'
of the scattering length $a_s$ out of the problem, leaving $k_F$ as
the only relevant length
scale.
In particular, the ground state energy $E_0$,
is expected
to be proportional to that of the non-interacting
gas $E_0^{free}$ \cite{baker99}, that is $E_0/E_0^{free}=\xi $
, or equivalently
\begin{equation}
\frac{E_0}{A}=\frac{3}{5}\frac{k_F^2}{2}\xi
\end{equation}
($\hbar=m=1$), $A$ being the number of particles.
The universal constant $\xi$ is of great interest and many attempts
have been made to derive it analytically or determine it experimentally.

Theoretical calculations suggest that $\xi$  is between 0.3 to 0.7.
For example, an early work based on different Pad\'{e} approximations
gives $\xi=0.326, 0.568$\cite{baker99}.
Diagrammatic approach gives 0.326 with Galitskii resummation\cite{hei01},
0.7 with ladder approximation\cite{hei01}, and
0.455 with a diagrammatic BCS-BEC crossover theory\cite{perali04}.
Other theoretical
approaches have also been used, including
$\epsilon$ expansion, which gives
$\xi$=0.475 in \cite{nishi06} and \cite{chen06},
and variational formalism, which gives 0.360 in \cite{hauss07}.
The four most recent experimental measurements
are listed in Table \ref{exp}.
Though the experimental results are consistent with each other,
the experimentally determined value of $\xi$ still falls
between relatively large error bars($\sim$10\%). By far the best
estimate on $\xi$ is considered to be that from
 Quantum Monte-Carlo
methods, giving $\xi=0.44(1)$\cite{carlson03} and 0.42(1)\cite{astra04}.

\begin{table}[here]
\begin{tabular}{clc}
$\xi$ & Authors& Ref.\\ \hline \hline
0.36(15)& Bourdel {\it el.al}&\cite{bourdel05}\\
0.51(4) & Kinast {\it et.al.} & \cite{kinast05}\\
0.46(5) & Partridge {\it et.al.} & \cite{part06}\\
$0.46^{+0.05}_{-0.12}$ & Stewart {\it et.al.}&\cite{stewart06}
\end{tabular}
\caption{Comparison of recent experimental values on $\xi$.\label{exp}}
\end{table}

Cold and dilute neutron matter is a special class of cold Fermi system
with great importance in astrophysics.
Its properties at resonance has attracted much interest recently
\cite{schwenk05,lee06}.
In this work we report  results from low-momentum ring diagram
calculations on the ground-state energy of neutron matter
at and near the unitary limit. As is  well-known, the $^1S_0$
channel of neutron matter has a fairly large
scattering length $a_s$ ($-18.97fm$), nonetheless, it is still
finite. Here, by adjusting
the interaction parameters of the CD-Bonn potential \cite{cdbonn},
we construct `tuned' neutron interactions with different $a_s$'s
such as  $-9.83fm$,  $-12070fm$ and $+21fm$ (which possesses a bound state).
 For a wide range of
neutron density, the case of $a_s=-12070fm$ can be considered
the same as the unitary limit, namely $a_s\rightarrow -\infty$. We shall
compute the ground state energy  of
neutron matter, with inter-neutron potentials being these
`tuned' CD-Bonn's, by two steps:
renormalization followed by ring summation.
We first renormalize neutron interactions
with a T-matrix equivalence renormalization method
\cite{bogner01,bogner02,coraggio02,schwenk02,bogner03,jdholt}, where the
high-momentum components beyond a decimation scale $\Lambda$
are integrated out. This
gives the corresponding low-momentum
interactions $V_{low-k}$'s with the scattering lengths being
preserved. Then, we calculate the ground state energy
 by summing the particle-particle-hole-hole
($pphh$) ring diagrams\cite{song87} to all orders. In
such ring summation, we employ a model space approach, namely,
the summation is carried out
within a model space characterized by $\{k\leq \Lambda\}$.

We shall closely examine how our results differ from similar
calculations with a different renormalized interaction -
the Brueckner $G$-matrix on which the
 Brueckner Hartree-Fock(BHF) method is based.
The BHF method has been widely used
for treating the strongly interacting nuclear many body problems
\cite{bethe,jeholt}. However, BHF is a lowest-order
reaction matrix ($G$-matrix)
theory and may be improved in several aspects.
To take care of the short range correlations, the ladder diagrams
of two particles interacting with the bare interaction are summed to all
orders in BHF. However, this method does not include
 diagrams representing hole-hole correlations such as diagram (iii) of Fig.1.
Note that this diagram has repeated  $(pphh)$
interactions as well as self-energy insertions to both hole and particle
lines. Another aspect of the traditional BHF is that it employs a discontinuous
 single-particle (s.p.) spectrum which has a gap at the Fermi surface $k_F$.
To improve upon these drawbacks,
 Song et al. \cite{song87} have formulated a $G$-matrix ring-diagram
method for nuclear matter, with which  the
$pphh$ ring diagrams such as diagrams (i) to (iii) of Fig.1 are summed
to all orders. This ring-diagram method
has been applied to nuclear matter and given satisfactory result
\cite{song87}. The $V_{low-k}$ ring diagram method used in this work
is highly similar to \cite{song87}'s ,
  except for one significant difference:
the  interaction used in the $G$-matrix ring diagram method
is energy dependent. (The Brueckner $G$-matrix is energy dependent,
as we shall later discuss.) This complicates
the calculation a lot. $V_{low-k}$ provides a cleaner
and simpler implementation on such all-order ring summation.

%siu: I added some in next paragraph

We shall first provide an outline of the ring-diagram
approach in section II. The derivation details of the low-momentum
interaction from the CD-Bonn potentials shall be followed
in section III.
Our major results from the $V_{low-k}$ ring diagram method
are in section IV. There we shall present our results for the ground-state
energy and ratio $E_0/E_0^{free}$ obtained with potentials of
various scattering lengths. A fixed-point criterion for determining the
decimation scale $\Lambda$ will be discussed.  There one can also find
a comparison
of data on the ground state energy obtained with two different methods-the
$V_{low-k}$ and the $G$-matrix ring diagram methods. We shall summarize
and discuss our work in the last section.

%------------------------------------------------------------------------
%Starting from a bare interaction, the
%above renormalization method
%can lead to a family of different Hermitian low-momentum interactions, depending
%on the tranformation methods \cite{jdholt} employed. Among them are
% the Okubo and Cholesky Hermitian low-momentum interactions.
%We shall use both of these interactions and discuss their results
%in the present work.

%----------------------------------------------------------------------
\section{Low-momentum ring diagrams}
In this section we describe how we calculate the ring diagrams
for the ground state energy shift $\Delta E_0$,
which is defined as the difference
$(E_0-E_0^{free})$ where $E_0$ is the true ground-state energy
 and $E_0^{free}$ is the corresponding quantity for the
non-interacting system. In the present work, we consider the $pphh$ ring
diagrams as shown in Fig. 1. We shall  calculate the all-order
sum, denoted as $\Delta E_0 ^{pp}$, of such diagrams. Our calculation
 is carried out
within a low-momentum model space $\{k\leq \Lambda\}$ and each vertex
of the diagrams is the renormalized effective interaction corresponding
to this model space. Two types of such interactions will be employed,
one being the energy-independent $V_{low-k}$ and the other being
the energy-dependent $G$-matrix interaction. Let us consider first
the former.
In this case, $\Delta E_0^{pp}$  can be written \cite{song87} as
\begin{eqnarray}
\Delta E_0^{pp}&=& \frac{-1}{2\pi i}\int _{-\infty}^{\infty}
        d\omega e^{i\omega 0^+}
    tr_{<\Lambda}[F(\omega)V_{low-k}\nonumber\\
   && +\frac{1}{2}(F(\omega)V_{low-k})^2
    +\frac{1}{3}(F(\omega)V_{low-k})^3+\cdots]
\end{eqnarray}
where $F$ is the free $pphh$ propagator
\begin{equation}
F_{ab}(\omega)=
\frac{\bar n_a \bar n_b}{\omega-(\epsilon_a+\epsilon_b)+i0^+}
-\frac{ n_a  n_b}{\omega-(\epsilon_a+\epsilon_b)-i0^+}
\end{equation}
with $n_a=1,~a\leq k_F; ~=0,~k>k_F$ and $\bar n_a=(1-n_a)$.

 We now introduce a strength parameter
$\lambda$ and a $\lambda$-dependent  Green function
$G^{pp}(\omega,\lambda)$ defined by
\begin{equation}
G^{pp}(\omega,\lambda)=F(\omega)
+\lambda F(\omega)V_{low-k}G^{pp}(\omega,\lambda).
\end{equation}
The energy shift then takes the following simple form
when expressed in terms of $G^{pp}$, namely
\begin{equation}
\Delta E_0^{pp}=\frac{-1}{2\pi i}\int_0^1 d\lambda
\int_{-\infty}^{\infty}e^{i\omega 0^+}tr_{<\Lambda}
[G^{pp}(\omega,\lambda) V_{low-k}]
\end{equation}
Using Lehmann's representation for $G^{pp}$, one can show
that
\begin{equation}\label{eng}
\Delta E^{pp}_0=\int_0^1 d\lambda
\Sigma_m\Sigma_{ijkl<\Lambda}Y_m(ij,\lambda)Y_m^*(kl,\lambda)
 \langle ij|V_{low-k}|kl \rangle,
\end{equation}
where the transition amplitudes $Y$ are given by the following
 RPA equation:
 \begin{eqnarray}
&&\sum _{ef}[(\epsilon_i+\epsilon_j)\delta_{ij,ef}+
\lambda(1-n_i-n_j)\langle ij|V_{low-k}|ef\rangle] \nonumber \\
&& \times Y_m(ef,\lambda)
=\omega_mY_m(ij,\lambda);~~(i,j,e,f)<\Lambda. \label{rpa}
\end{eqnarray}
The index $m$ denotes states dominated by hole-hole components,
namely, states that satisfy
$\langle Y_m|\frac{1}{Q}|Y_m\rangle=-1$ and $Q(i,j)=(1-n_i-n_j)$.
We have used  the HF s.p. spectrum given by $V_{low-k}$, namely
\begin{equation}\label{sp}
 \epsilon_k = \hbar^2k^2/2m +\sum _{h<k_F}\langle kh|V_{low-k}|kh\rangle
\end{equation}
for both holes and particles with $k\leq \Lambda$. Thus the propagators
of the diagrams as shown in Fig. 1 all include HF insertions to all
orders. The above spectrum is continuous up to $\Lambda$.

The above ring-diagram method is
  a renormalization group approach for a momentum model
space defined by a momentum   boundary $\Lambda$, and the space
with momentum greater than $\Lambda$ is integrated out. The resulting
effective interaction for the model space  is $V_{low-k}$
which is energy independent. This renormalization procedure
can, however, also lead to a model-space effective interaction which
is energy dependent. The $G$-matrix ring-diagram method
of \cite{song87} is of the latter approach. Formally, these
two approaches should be the same. In the present work
we shall carry out ring-diagram calculations using both approaches;
it would be of interest to compare the results of these two different
approaches.

%-------------------------------------------------------------------------
In the following, let us briefly describe the $G$-matrix ring diagram
method \cite{song87}. Here each vertex of Fig. 1 is
 a model-space $G$-matrix interaction,
to be denoted as $G^M$.  It is defined by
\begin{equation}
G^M_{ijkl}(\omega)=V_{ijkl}+\sum_{rs}V_{ijrs}\frac{Q^M(rs)}
{\omega-k_r^2-k_s^2+i0^+}G^M_{rskl}(\omega)
\end{equation}
where $k_r^2$ stands for the kinetic energy $\hbar^2k_r^2/2m$
and similarly for $k_s^2$. The Pauli projection operator $Q^M$
is to assure  the intermediate states being outside $\Lambda$ and
$k_F$, namely it is defined by
\begin{eqnarray}
Q^M(rs)&=&1, if~ max(k_r,k_s)>\Lambda~ and~ min(k_r,k_s)<k_F \nonumber \\
       &=&0, otherwise.
\end{eqnarray}
In the above $k_F<\Lambda$.  In Ref.\cite{song87} $\Lambda$ is chosen
to be $\sim 3 fm^{-1}$.
Note that the above $G^M$  is energy dependent, namely
it is dependent on the energy variable $\omega$. However, $\omega$
is not a free parameter; it is to be determined in a self-consistent way.
For example, the model-space s.p. spectrum is given by
the following self-consistent equations:
\begin{equation}
\epsilon _a=\frac{\hbar^2k_a^2}{2m}+\langle a|U|a \rangle;
\end{equation}
\begin{eqnarray}
\langle a |U|a \rangle &=&\sum_{h\leq k_F}\langle a,h|G^M(\omega=\epsilon_a
+\epsilon_h)|a,h\rangle, ~a<\Lambda  \nonumber \\
    &=&0, ~ otherwise.
\end{eqnarray}
In the above $U$ is the s.p. potential and $\epsilon$  the
model-space s.p. energy which is determined self-consistently
with the  energy variable of $G^M$.  Note that
this s.p. spectrum does not have a gap at $k_F$; it is a continuous one
up to $\Lambda$. When choosing $\Lambda$=$k_F$
the above is the same as  the self-consistent BHF s.p. spectrum.

 When calculating the ring diagrams  using
$G^M$, its energy variable is also determined self-consistently.
In terms of $G^M$,
the all-order sum of
the $pphh$ ring diagrams is \cite{song87}
\begin{equation}
\Delta E^{pp}_0=\int_0^1 d\lambda
\sum_{m}\sum_{ijkl(<\Lambda)}Y_m(ij,\lambda)Y_m^*(kl,\lambda)
G^M_{kl,ij}(\omega_m^-)
\end{equation}
where the transition amplitudes $Y_m$ and eigenvalues $\omega_m^-$ are given
by the following self-consistent RPA equation:
 \begin{eqnarray}
&&\sum _{ef}[(\epsilon_i+\epsilon_j)\delta_{ij,ef}+
\lambda(1-n_i-n_j)L_{ij,ef}(\omega)]Y_m(ef,\lambda)\nonumber\\
&&=\mu_m(\omega,\lambda)Y_m(ij,\lambda);~~(i,j,e,f)<\Lambda.
\end{eqnarray}
The index $m$ denotes states dominated by hole-hole components.
The vertex function $L$ is obtained from 2- and 1-body diagrams
first order in $G^M$ \cite{song87}. The above equation is solved with the
self-consistent condition that the energy variable of $L$ is equal
to the eigenvalue, namely
\begin{equation}
\omega=\mu_m(\omega,\lambda)\equiv \omega_m^-(\lambda).
\end{equation}

Comparing with the $V_{low-k}$ ring diagram calculation described earlier,
the above $G$-matrix calculation is clearly more complicated.
Because of the energy dependence of the interaction $G^M$, the above
equations have to be solved self-consistently both for the s.p. spectrum
and for the RPA equations. To attain this self consistency, it is necessary
to use iteration methods and this procedure
is often  numerically involved. In contrast, ring-diagram calculation
using the energy-independent interaction $V_{low-k}$ is indeed
much simpler. As mentioned earlier, we shall carry out ring-diagram
calculations using both methods.

\begin{figure}[here]
%to scale the figure ,use \scalebox[ratio]{}
\scalebox{0.5}{
\includegraphics{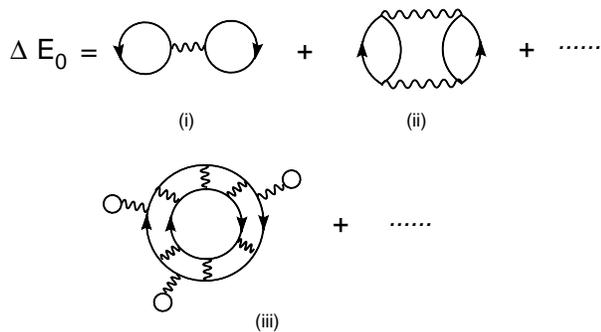}
}%scalebox
\caption{\label{energy}$pphh$ ring-diagram summation in the
calculation of the ground state energy shift.}
\end{figure}

\section{$V_{low-k}$ with infinite scattering length }
To carry out the above ring-diagram calculation, we need
the low-momentum potential $V_{low-k}$. Since we are interested
at neutron matter at and near the unitary limit (infinite scattering length),
we should have  $V_{low-k}$'s of definite scattering lengths, including
$\pm \infty$, so that the dependence of our results on scattering lengths
can be investigated.
In the present work, we have chosen a two-step procedure to construct
such potentials so that the resulting potentials are close to realistic neutron
potentials. We first construct bare potentials $V^a$ based on a realistic
nucleon-nucleon potential; these potentials are tuned so that they have
definite scattering lengths. Renormalized low-momentum potentials
$V_{low-k}^a$ are then obtained from $V^a$ using a renormalization procedure
which preserves the scattering length.

We start from the high-precision
CD-Bonn \cite{cdbonn} nucleon-nucleon potential. For this potential,
the scattering length of the $^1S_0$ channel is already fairly
large (-18.97 $fm$), and  it is found  to depend rather sensitively
on the interaction
parameters. Thus by slightly tuning
the interaction parameters of the CD-Bonn potential, we have obtained
a family of $^1S_0$ neutron potentials of definite scattering lengths.
We shall denote them as $V^a$. Our tuning procedure will be discussed
in section IV(A).

Recently there have been a number of studies on the low-momentum
nucleon-nucleon potential $V_{low-k}$
\cite{bogner01,bogner02,coraggio02,schwenk02,bogner03,jdholt}.
$V_{low-k}$ is obtained from a bare nucleon-nucleon potential
by integrating out
the high-momentum components, under the restriction that the
deuteron binding energy and the low-energy phase-shifts are
preserved. The $V_{low-k}$ obtained from different realistic
potentials ( CD-Bonn \cite{cdbonn}, Argonne \cite{argonne}
, Nijmegen \cite{nijmegen} and Idaho \cite{chiralvnn})
 all flow to a unique potential when the cut-off
momentum is lowered to around $2fm^{-1}$.
The above $V_{low-k}$ is obtained using a T-matrix equivalence
renormalization procedure
\cite{bogner01,bogner02,coraggio02,schwenk02,bogner03,jdholt}.
Since this procedure  preserves the half-on-shell
T-matrix,  it of course preserves the scattering length.
Thus this procedure is suitable  for constructing $V_{low-k}^a$,
the low-momentum interaction with definite scattering length.
Using this procedure, we start from the $T$-matrix equation
\begin{equation}
  T(k',k,k^2)
= V^a(k',k)
 + \int _0 ^{\infty} q^2 dq  \frac{V^a(k',q)T(q,k,k^2 )} {k^2-q^2 +i0^+ }  ,
\end{equation}
where $V^a$ is a modified CD-Bonn potential of scattering length $a$.
Notice that in the above the intermediate state momentum $q$ is
integrated from 0 to $\infty$.
We then define an effective low-momentum T-matrix by
\begin{eqnarray}
  T_{low-k }(p',p,p^2) &=& V^a_{low-k }(p',p) \nonumber \\  \nonumber
 &+& \int _0 ^{\Lambda} q^2 dq  \frac{V^a_{low-k }(p',q) T_{low-k} (q,p,p^2)}
{p^2-q^2 +i0^+ },\\
\end{eqnarray}
where the intermediate state momentum is integrated from
0 to $\Lambda$, the momentum space cut-off.
We require the above T-matrices to satisfy the condition
\begin{equation}
 T(p',p,p^2 ) = T_{low-k }(p',p, p^2 ) ;~( p',p) \leq \Lambda.
\end{equation}
The above equations define  the effective low momentum interaction
 $V_{low-k}^a$. The iteration method of Lee-Suzuki-Andreozzi
\cite{suzuki80,andre96}  has been used in calculating $V_{low-k}^a$
from the above T-matrix equivalence equations.
From now on, we shall denote
$V_{low-k}^a$ simply as $V_{low-k}$.

%Note that the above $V_{low-k}^a$  is not directly used in our ring
%diagram calculations, because the effective interaction given
%by the above iteration method is  not Hermitian.
%A further transformation is needed to make it Hermitian.
%In the present work we have used both the Okubo and Cholesky transformations
%\cite{jdholt} to obtain the Hermitian low-momentum interactions
%used in our ring-diagram calculations.

\section{Results}

\subsection{Low-momentum interactions and scattering lengths}
To study neutron matter at the unitary limit, we first need a realistic
neutron-neutron interaction that would lead to a huge $^1S_0$
scattering length $a_s$, and a small effective range $r_e$. We obtain such
interaction by `tuning' the meson mass $m_\sigma$ in the usual CD-Bonn
potential. The exchange of a lighter meson
generates a stronger attraction, therefore making the scattering length $a_s$
more negative until a bound state is formed. As one `tunes' across
the bound state, $a_s$ will pass from $-\infty$ to $+\infty$ , eventually
become less and less positive.
In this work,
 this $m_\sigma$ `tuning' is taken as a manual adjustment in the strength of the
 neutron-neutron potential.
Of great interest is that this `tuning' may naturally come
from the density-dependence of the nucleon-nucleon potential
via the mechanism of Brown-Rho (BR) scaling\cite{brown91,brown04,rapp99}, which
suggests the in-medium
meson masses should {\it decrease}.

At normal nuclear matter density, the meson masses of $\rho$, $\omega$ and
$\sigma$ are all expected to decrease by about $15\%$ \cite{rapp99}
compared to their masses in free space. This decrease will enhance not only
the attraction from $\sigma$ but also the repulsion from $\rho$ and $\omega$.
As a preliminary study,  we shall tune only $m_{\sigma}$ in the present work.
To compensate for the repulsive effect from $\rho$ and $\omega$ (which are
not tuned in the present work), we
shall only tune $m_{\sigma}$ slightly, namely a few percent.
We shall consider that the above BR scaling is compatible with neutron
matter of moderate density ($k_F \sim 1fm^{-1}$). In a future publication,
we plan to carry out
further studies, including the tuning of $\rho$- and $\omega$-meson
masses.

\begin{table}
\begin{tabular}{|c|c|r|r|}
\hline
name & $m_\sigma(MeV)$& $a_s(fm)$ & $ r_e(fm) $\\
  \hline \hline
original CD-Bonn&      452           & -18.97 & 2.82\\ \hline
CD-Bonn-10     &   460      & -9.827 & 3.11\\ \hline
CD-Bonn-42     &   447      & -42.52 & 2.66 \\ \hline
CD-Bonn-$\infty$     & 442.85          & -12070.00 & 2.54\\ \hline
CD-Bonn+$\infty$     & 442.80          & +5121.00   & 2.54\\ \hline
CD-Bonn+21  & 434        & +21.01 & 2.31\\
 \hline
\end{tabular}
\caption{$m_\sigma$ in the original CD-Bonn potential is tuned
to give neutron-neutron potentials with different scattering lengths.\label{meson}}
\end{table}
Various `tuned' CD-Bonn potentials are listed in Table \ref{meson}.
From there one can see the sensitivity of the scattering length
 to the change in $m_\sigma$.
At $m_\sigma\approx442MeV$, namely a 2.4$\%$ decrease from the original,
$a_s\approx-12000fm$.
%Notice that in such case, the potential also has
%a small effective range $r_e=2.539fm$.
 Notice that the effective ranges
for the CD-Bonn potentials are larger than the actual ranges of them.
For example, $r_e$ for the original CD-Bonn potential is $2.82fm$,
considerably larger than the range of one-pion exchange. Within the range
of Fermi momenta from $0.8fm^{-1}$ to $1.5fm^{-1}$ that we use
in our computation below,  $a_s\approx-12000fm$ is obviously
enormous compared to
any length scale in the system, thus we expect the neutron matter to be
at the unitary limit, i.e., no different from the limiting case
$a_s=-\infty$. For convenience, we name such potential CD-Bonn-$\infty$.

\begin{figure}[here]
%to scale the figure ,use \scalebox[ratio]{}
\scalebox{0.5}{
\includegraphics{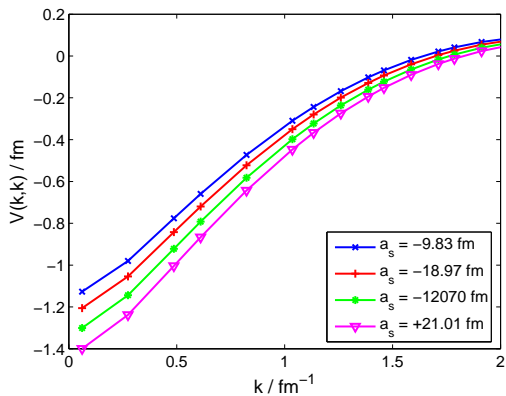}
}%scalebox
\caption{\label{compare_cdbs}Diagonal matrix elements of
$V_{low-k}$ constructed from CD-Bonn potentials with different
scattering lengths. $\Lambda=2.4fm^{-1}$ is used in all cases. }
\end{figure}

Following the renormalization procedures as already described in Section III, we
obtain the low-momentum potential $V_{low-k}$'s for several CD-Bonn potentials
 listed above. A comparison of the diagonal matrix elements
in the $V_{low-k}$'s (with a fixed cut-off momentum $\Lambda$)
 is shown in Figure \ref{compare_cdbs}. It is of interest that the strength
of $V_{low-k}$ only changes weakly with the scattering length. For example,
it changes by merely about $10\%$ from $a_s=-18.97fm$ to $-12070fm$.

\subsection{Ground-state energy and \\ the universal constant $\xi$}
Here we shall present our major results, namely the ground state energies
$E_0$ of neutron matter at and close to the unitary limit from the summation of
low-momentum ring diagrams to all orders.
Following the potential renormalization procedure described in section III,
we first calculate $V_{low-k}$ for  certain chosen values for the decimation
scale $\Lambda$.
Then  the all-order sum of the $pphh$ ring diagrams are calculated
using the above $V_{low-k}$.
As introduced in Section II, the calculation details in the summation of
$pphh$ ring diagrams can be found in Ref.\cite{song87}.
 How to choose the decimation scale $\Lambda$ is clearly an
important step in our calculation, and in the present work we shall use
a stable-point, or `fixed-point', criterion
in deciding $\Lambda$. Before discussing this criterion, let us first
present some of our results for
the ground-state energy per particle
$(E_0/A)$.  In Fig. 3 we present such results for four $a_s$ values,
calculated with $\Lambda$s determined by the above criterion.
(The details of this determination will be described a little later.)
As shown by the figure,  we see that $E_0/A$ does not change
strongly with $a_s$. The ratios
 $\xi=E_0/E_0^{free}$ are then readily obtained, as shown in Fig.4.
It is of interest that the ratios for the four $a_s$ cases are all
weakly dependent on $k_F$. To help understand this behavior,
 we plot in Fig.5 the potential energy per particle $PE/A$ (namely
$\Delta E_0^{pp}/A$ of Eq.(6)) versus $k_F^2$, for the same four
$a_s$ cases.
It is rather impressive that  they  all appear to be straight lines.
 We have fitted the `lines' in the figure to
the equation $PE/A=(\hbar^2/m)\left(\beta k_F^2+\gamma\right)$:
We have found $(\beta,\gamma)$ =(-0.1370, 0.0002),
(-0.1498, -0.0008), (-0.1649, -0.0035) and (-0.1797, -0.0082)
respectively for $a_s$ = $-9.87fm$, $-18.97fm$, $-12070fm$ and $+21.0fm$.
The rms deviation for the above fitting are all very small
(all less than 0.0013),
confirming that they are indeed very close to straight lines.
The above results are of interest, and are consistent with those
 shown in Fig. 4. In fact the ratios of Fig.4
are determined by the `slopes' of these `lines'.

%Fig.3  $E_0/N$
\begin{figure}[here]
%to scale the figure ,use \scalebox[ratio]{}
\scalebox{0.5}{
\includegraphics{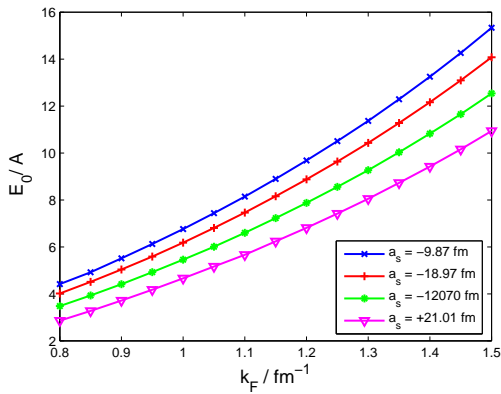}
}%scalebox
\caption{\label{compare_cdbs_be} Ground state energy per particle,
$E_0/A$, of neutron matter with various tuned CD-Bonn potentials, computed
from the summation of low-momentum $pphh$ ring diagrams.
Only $^1S_0$ contribution is included.  }
\end{figure}

%Fig.4  ratios
\begin{figure}[here]
%to scale the figure ,use \scalebox[ratio]{}
\scalebox{0.6}{
\includegraphics{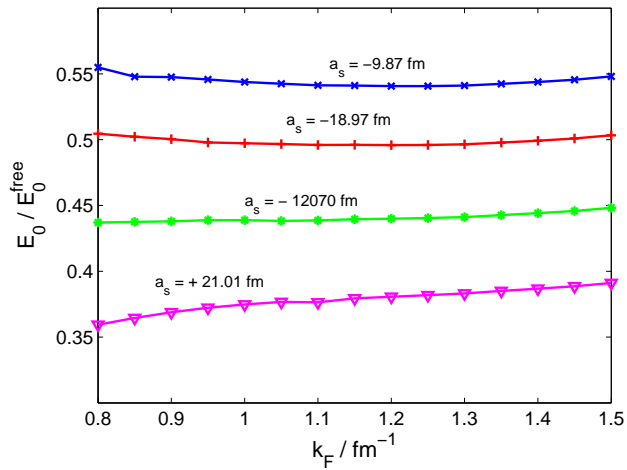}
}%scalebox
\caption{\label{beke_kf_allcdbs}The ratio $E_0/E_0^{free}$ as a function of
Fermi momentum $k_F$ for the various CD-Bonn potentials listed in Table \ref{meson}.
The data with CD-Boon-$\infty$($a_s=-12070fm$)
 indicates that $E_0/E_0^{free}$ is a constant
of $0.443\pm0.006$ over the range of $k_F$ as shown.}
\end{figure}

%Fig.5  PE/N
\begin{figure}[here]
%to scale the figure ,use \scalebox[ratio]{}
\scalebox{0.6}{
\includegraphics{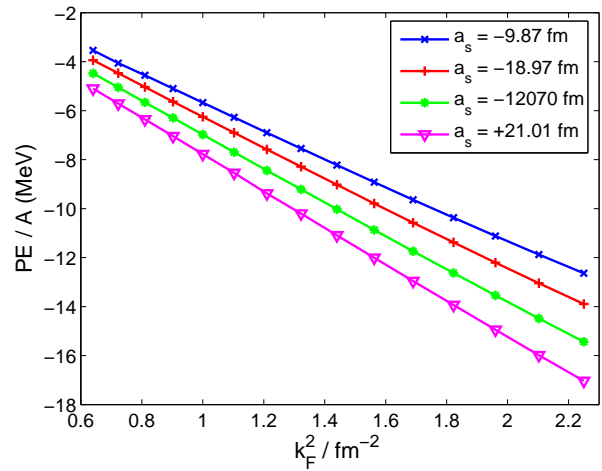}
}%scalebox
\caption{\label{pe_kfsq_allcdbs}Potential energy per particle, $PE/A$,
of neutron matter with various tuned CD-Bonn potentials, computed
from the summation of low-momentum $pphh$ ring diagrams.
Only $^1S_0$ contribution is included.}
% The straight lines are fitted to
%the equation $PE/A=(\hbar^2/m)\left(\beta k_F^2+\gamma\right)$,
%and we found $\beta=-0.1370$, $\gamma=0.0002$ for $a_s=-9.87fm$;
%$\beta=-0.1498$, $\gamma=-0.0008$ for $a_s=-18.97fm$;
%$\beta=-0.1649$, $\gamma=-0.0035$ for $a_s=-12070fm$;
%$\beta=-0.1797$, $\gamma=-0.0082$ for $a_s=+21.01fm$.}
\end{figure}

Before further discussing our results, let us now address the question
of how to determine the decimation scale $\Lambda$. There are basically
two considerations: The first one concerns the experimental NN
scattering phase shifts on which realistic NN potentials
are based. The second is about the dependence of our results on $\Lambda$.
Realistic NN potentials \cite{cdbonn,argonne,nijmegen,chiralvnn} are
constructed to reproduce the experimental NN phase shifts up to
$E_{lab}\approx 300MeV$. This suggests that $\Lambda$ is about
$2fm^{-1}$, as beyond this scale NN potential models are
not experimentally constrained and are thus rather
uncertain (model dependent) \cite{bogner03}.

We now turn to  the dependence of our results on $\Lambda$.
As described in Section II,
$V_{low-k}$ is used in the determination of the H.F. single
particle spectrum (see Eq.\ref{sp}),  the transition amplitudes
$Y$ in the RPA
equation (see Eq.\ref{rpa}), and finally, the ground state
energy $E_0$ (see Eq. \ref{eng}). Intuitively, $E_0$
 should exhibit a non-trivial
$\Lambda$-dependence. For various Fermi-momenta,
this dependence is studied and is found to be remarkably mild.

As an example, let us present in Fig. 6  our results obtained
with the potential CD-Bonn-$\infty$.
For $\Lambda=(2.0-2.6)fm^{-1}$, it is seen that $\xi$ varies actually
by a rather small amount (note that the range of our plot is
from 0.438 to 0.444). Furthermore the $\Lambda$ dependence of $\xi$ shows up
as a curve with a minimum.
The final choice of $\Lambda$ is based on the criterion that $E_0$
should be stable against changes in $\Lambda$.
As shown in the figure, an obvious stable-point, or fixed-point,
defined by $dE_0(\Lambda)/d\Lambda=0$,
is found at about $2.3fm^{-1}$. Thus we have used $\Lambda=2.3 fm^{-1}$
 for CD-Bonn-$\infty$. We found that the position of the fixed point
is almost the same for the different Fermi-momenta in the range
$(0.8-1.5)fm^{-1}$.
The same procedure is done on the original CD-Bonn,
 and other tuned potentials. The fixed points, also with an negligible
 dependence on $k_F$, are found to be $2.15fm^{-1}$, $2.25fm^{-1}$ and
$2.4fm^{-1}$ respectively
 for CD-Bonn potentials of scattering lengths
 $-9.8fm$, $-18.9fm$ (the original CD-Bonn), and $+21.01fm$.
The above fixed-point $\Lambda$'s have been used for the results
presented in Figs. 3-5.

\begin{figure}[here]
%to scale the figure ,use \scalebox[ratio]{}
\scalebox{0.5}{
\includegraphics{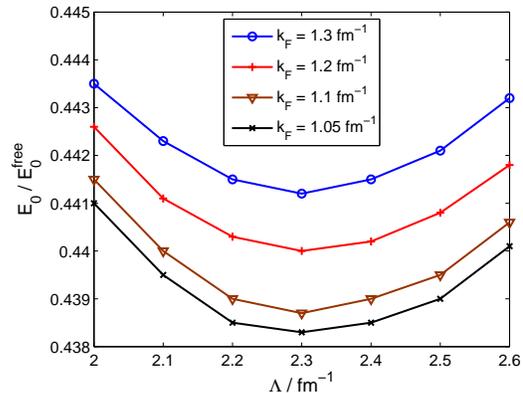}
}%scalebox
\caption{\label{beke_lam}Determination of the fixed point where $dE_0/d\Lambda=0$ for
CD-Bonn-$\infty$.  }
\end{figure}

Of great significance
is the ratio of the ground state energy to that of the non-interacting case,
namely $E_0/E_0^{free}$. At the unitary limit, it is expected to be an
universal constant, named $\xi$. This constant is of great importance as it
determines the equation of state of all low-density cold Fermi gas.
At the unitary limit, our data on $E_0/E_0^{free}$
all lie within a narrow window from 0.437 to 0.448.
%which suggests $\xi=0.4425\pm0.0055$.
 Such result confirms a
universality over Fermion density in a wide range
$(1.73-11.40)\times10^{-2}fm^{-3}$.
Most importantly, the numerical value of $\xi$ is remarkably close to that from
Monte Carlo methods, which by far is believed to be the best estimate.
Astra {\it et. al.} obtained 0.42(1) based on a square well
potential and particle density
$nR_0^{3}=10^{-6}$ (where $R_0$ is the potential range).
Carlson {\it et. al.} obtained 0.44(1) based on a `cosh potential',
and particle density
$n\mu^{-3}=0.020$ (where $2/\mu$
is the effective range). In our case, $n\Lambda^{-3}=(1.4-9.4)\times 10^{-3}$
(where $\Lambda=2.3fm^{-1}$ is the decimation scale in the renormalization).
These works, including ours,
 employ very different interactions and
various particle densities. Still, the value of $\xi$ agrees incredibly well.

In Figure \ref{beke_kf_allcdbs} we contrast the data
from CD-Bonn-$\infty$  with that from the
original CD-Bonn and other tuned potentials.
Even though the $^1S_0$  scattering length in the original CD-Bonn
is already fairly
large ($a_s=-18.97fm$) , still the equation of state, as predicted from the
ratio $E_0/E_0^{free}$, has significant difference from the unitary limit.
As seen in our data with CD-Bonn-$\infty$ potential,
at the unitary limit the ratio $E_0/E_0^{free}=0.44$
is practically independent of  the underlying neutron density $n$.
%Notice that this universality holds over quite a wide range of $n$,
%from $1.73\times10^{-2}fm^{-3}$ to $11.40\times10^{-2}fm^{-3}$ as
%shown in Figure \ref{beke_kf_allcdbs}.
% This disappearance of
%density dependence is of great interest.

\subsection{Comparison with G-matrix results}
As discussed in section II, our ring-diagram calculations are based
on a model space framework. A model-space is defined by momentum
$\{k\leq\Lambda\}$ where $\Lambda$ is the decimation scale.
The space with $k>\Lambda$ is integrated out, resulting in a model-space
effective interaction $V_{eff}$. We have used so far the energy-independent
$V_{low-k}$ for $V_{eff}$. Alternatively, on can also use the
energy-dependent $G^M$-matrix (of section II) as $V_{eff}$. These two
approaches are formally equivalent. We have carried out calculations
 to check this equivalence.

\begin{figure}[here]
%to scale the figure ,use \scalebox[ratio]{}
\scalebox{0.6}{
\includegraphics{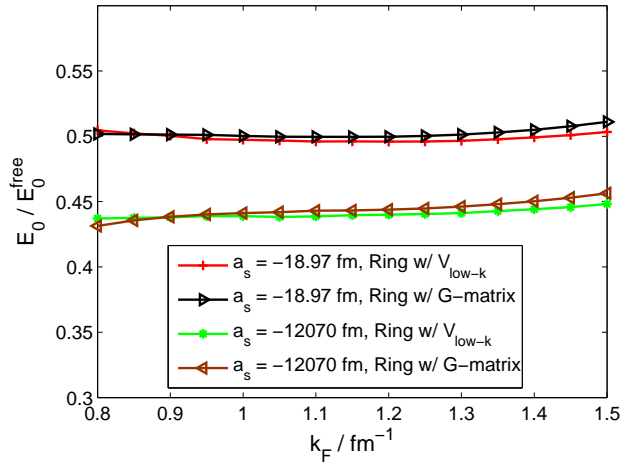}
}%scalebox
\caption{\label{beke_kf_cdbinf}The ratio $E_0/E_0^{free}$ for the
potentials CD-Bonn-$\infty$ and CD-Bonn(-19.87) computed with two methods.
Ring w/ $G$-mat: $pphh$ ring diagrams summation with Brueckner $G^M$-matrix.
$\Lambda=2.3fm^{-1}$ is used, computation is fully
self-consistent. Ring w/ $V_{low-k}$: $pphh$
ring diagrams summation with $V_{low-k}$, fixed point is at $\Lambda=2.3$.}
\end{figure}

We have   repeated
the ring diagram summation with
the energy-independent $V_{low-k}$ replaced by the energy-dependent
model-space Brueckner $G^M$-matrix, and
carry out a fully self-consistent computation in summing up the
$pphh$ ring diagrams.
The exact procedures in Ref.\cite{song87} are followed (section II).
Ring diagrams within a model space up to
a cut-off momentum $\Lambda$ is summed to all orders.
We found that the ground state energy is rather insensitive
to the choice of $\Lambda$.
See Figure \ref{beke_kf_cdbinf} for the  data
of CD-Bonn-$\infty$ and CD-Bonn(-18.97), done with
$\Lambda=2.3fm^{-1},~2.25fm^{-1}$ respectively. As illustrated, the two
methods, namely, ring diagram summation with $V_{low-k}$ and
that with $G^M$-matrix, are fully consistent. This is a remarkable
and reassuring result, as the calculational procedures of them are
vastly different.
For the $G^M$ case, the s.p. spectrum, the RPA amplitudes $Y$ and
energies $\omega^-_m$ are all calculated self-consistently,
while for the $V_{low-k}$ case no such self-consistent procedures
are  needed. Clearly the $V_{low-k}$ ring-diagram method is more
desirable.

%It is satisfactory that the results from  the above two
%approaches agree so well..

%has also made us feel more confident over our results.

%For comparison, we have also carried out traditional self-consitent
%BHF calculations \cite{bethe,kuomavinh86,jeholt}. For the potential
%CD-Bonn-$\infty$, the results for $\xi$ are generally larger than
%the corresponding ring results. Also they vary appreciably with
%$k_F$. For example the BHF results for $\xi$ are *** and ***
%for $k_F$ = $0.8$ and $1.0fm^{-1}$ respectively; the corresponding
%ring results being *** and ***. Hole-hole correlations are not
%included in BHF, and their importance is indicated by the above
%difference.

%We have even more confidence over our result as we repeat
%We would like to give a remark regarding the fact that
%different transformations can be
%used in the construction of a Hermitian $V_{low-k}$
%in the renormalization process\cite{jdholt}.
%These different transformations result in very similar $V_{low-k}$'s.
%The Cholesky transformation is used in
%the data shown in Figure \ref{beke_lam} and
%\ref{beke_kf_allcdbs}. We
%have checked that the fixed points from Okubo
%transformation are the same. The universal
%constant $\xi$ from Okubo is $0.444\pm0.007$, highly
%consistent with Cholesky's.

%\caption{\label{beke_kf_cdbinf}The ratio $E_0/E_0^{free}$ for the
%potential CD-Bonn-$\infty$ computed with various methods.
%HF:Simple Hartree-Fock computation with the use of $V_{low-k}$. A
%fixed decimation scale $\Lambda=2.3fm^{-1}$ is used in all the data.
% BHF: A fully self-consistent Brueckner-Hartree-Fock computation.
%Ring w/ $G$-mat: $pphh$ ring diagrams summation with Brueckner $G$-matrix.
%$\Lambda=2.4fm^{-1}$ is used, computation is fully
%self-consistent. Ring w/ $V_{low-k}$: $pphh$
%ring diagrams summation with $V_{low-k}$, fixed point is at $\Lambda=2.3$.}
%\end{figure}

%\begin{figure}[here]
%to scale the figure ,use \scalebox[ratio]{}
%\scalebox{0.6}{
%\includegraphics{pe_kfsq_cdbinf.eps}
%}%scalebox
%\caption{\label{pe_kfsq_cdbinf}CD-Bonn-$\infty$ data indicates that
%the potential energy per particle is simply proportional to $k_F^2$
%at the unitary limit. }
%\end{figure}

\subsection{Schematic effective interaction at unitary limit}
At the unitary limit, the simple equation of state $E_0=\xi E_0^{free}$
in neutron matter suggests a very counter-intuitive nature in the
underlying system: strongly interacting fermions essentially can be
described by a non-interacting picture with an effective mass. This
unexpected `simplicity' can best be captured by a schematic interaction.
To illustrate this, let us consider neutron matter confined
in a closed Fermi sea $|\Phi_0(k_F)\rangle$. In other words, we consider
neutron matter in a one-dimensional model space. We denote the
effective interaction for this model space as $V_{FS}$. Then the potential
energy per particle  is
\begin{eqnarray}
\frac{PE}{A}&=&
\langle\Phi_0(k_F)|V_{FS}|\Phi_0(k_f)\rangle/A \nonumber \\
   &=&\frac{8}{\pi}
\int_0^{k_F}\left(1-\frac{3k}{2k_F}+\frac{k^3}{2k_F^3}\right)
\langle k|V_{FS}|k\rangle
k^2 \mathrm{d}k
\end{eqnarray}
where $k$ is the relative momentum.

% We want the above $PE/A$ to reproduce the ring results of Fig.5.
%As discussed in section IV.B, the results obtained there are nearly
%proportional to $k_F^2$, namely
%\begin{equation}
%\frac{PE}{A}\simeq \eta k_F^2.
%\end{equation}
%For CD-Bonn-$\infty$, $\eta$=-.1649 which is close to 1/6. To reproduce this
%result impose a strong costraint on $V_{FS}$. Let us assume a power series
%expansion $\langle k|V_{FS}|k \rangle=\Sigma _n C_n k^n$, n=0,1,2...
%To have $PE/A$ equal to $\eta k_F^2$, only $C_0$ is non-vanishing
%and is inversely proportional to $k_f$ at the unitary limit of
%infinite scattering length. $V_{FS}$ is a contact interaction.
%This leads us to propose a schematic effective
%interaction
  Suppose we take $V_{FS}$ as a contact effective interaction
\begin{equation}
V_{FS}=\frac{1}{\frac{S}{a_s}-\frac{2}{\pi}k_F}
\end{equation}
($\hbar=m=1$) where $S$ is a positive  parameter with
 $S<<|a_s|$. When $S$=1 and $k_F$ replaced
by $\Lambda$,  $V_{FS}$ is the same
 the effective interaction for the pion-less effective field theory
 \cite{bogner03,schafer05}.
Substituting the above into Eq.(19) gives
\begin{equation}
\xi=1+\frac{5}{9}\frac{1}{\frac{\pi}{2}\frac{S}{a_sk_F}-1}.
\end{equation}

 At the unitary limit (infinite $a_s$), the above gives $\xi$=4/9,
independent of $k_F$, which is practically the same as the
result for $\xi$(-12070) of Fig. 4.
 The above also gives $\xi$ for finite $a_s$.
At the unitary limit, we expect $V_{FS}$ to be unique.
For finite $a_s$ (away from the unitary limit), it is not expected to
be unique and the parameter $S$ is expected to depend on the underlying
potential. As shown in Fig. 4, we have calculated $\xi$ using
the CD-Bonn potentials of finite scattering lengths.
%The results form approximately
%four equally-spaced parallel lines, with
%their spacing from the $a_s=-12070$ one being approximately
%proportional to $1/a_s$.
These results can also be  qualitatively described by the
above equation.
 For instance, for $S=1.25$ and $k_F=1.0$, the
above equation  gives $\xi$= 0.54, 0.50 and 0.39 respectively
for $a_s$=$-9.87fm$, $-18.97fm$ and $+21.01fm$. In short, certain main features
of our results obtained from  ring-diagram calculations with
the CD-Bonn potentials can be qualitatively
reproduced by the above simple contact effective interaction.

%The potential energy is found to be
%start by showing the potential energy per particle
%$(PE/A)$ of neutron matter at the unitary limit. In Figure \ref{pe_kfsq_cdbinf},
%the beautiful linear relation implies
%where $\eta$ is just a constant extremely close to 1/6. Based on this result
%we propose the following
% schematic effective interaction close to the unitary limit: We consider
%At the unitary limit, $a_s^{-1}=0$, we recover a linear relation between
%potential energy per particle and $k_F^2$.
%
%Our schematic model suggests the proportional constant
% $\eta$ exactly equals 1/6, or equivalently, $\xi$ exactly equals 4/9. We
%believe the simple structure in $V_{eff}$ can be understood as the result
%of a renormalization on the bare potential.  More
%work along this line of thought will be reported in the future.

\section{Summary }
In conclusion, we have carried out a detailed study on
neutron matter at and close to the unitary limit with a
low-momentum ring diagram approach. By slightly tuning the realistic
CD-Bonn potential, we have obtained $^1S_0$ neutron potentials
of specific scattering lengths, in particular
 the CD-Bonn-$\infty$ one with $a_s$ of $-12070 fm$. By integrating
out their momentum components beyond a decimation scale $\Lambda$, we obtain
renormalized low-momentum interactions $V_{low-k}$  of the same specific
scattering lengths. The ground state energy $E_0$ of neutron matter
are then calculated by summing up the $pphh$ ring diagrams to all
orders within the model space $\{k<\Lambda\}$. A fixed-point criterion
is used to determine the decimation scale $\Lambda$. We have carried out
ring-diagram calculations using two types of renormalized interactions,
the energy-independent $V_{low-k}$ and the energy-dependent $G$-matrix,
with results given by them being nearly identical.
The $V_{low-k}$ ring-diagram method has a simpler formalism and is also
more suitable for numerical calculation.
For the CD-Bonn-$\infty$ potential, the ratio $E_0/E_0^{free}$
 is found to be very near a universal constant of 0.44 over the neutron
density range
$(1.73-11.40)\times10^{-2}fm^{-3}$. Our result agrees
well with the recent experimental
measurement and Monte-Carlo computation on cold Fermi gas at the unitary limit.

\vskip 1cm
{\bf Acknowledgement} We thank G.E. Brown, E. Shuryak, T. Bergmann and
A. Schwenk for many helpful discussions. This work is supported in part
by U.S. Department of Energy  under grant DF-FG02-88ER40388, and by
 the U.S.
National Science Foundation under Grant PHY-0099444.

\vskip 1cm

%--------------------------------------------------------------------------------
%----------------------------------------------------------------------------------

\end{document}